\documentclass[3p,times]{elsarticle}
\usepackage{epsfig,mathptmx,times,amsmath,amssymb,color,flushend,gensymb,subfigure, ecrc,fancyhdr,amssymb,booktabs}
\usepackage[figuresright]{rotating}
\usepackage{framed,color,soul}
\definecolor{shadecolor}{rgb}{1,0.8,0.3}

\volume{24(8)}
\firstpage{1}
\journalname{Mechatronics}
\runauth{\textbf{PREPRINT VERSION:} E. Kayacan et al.}
\jid{Mechatronics}
\jnltitlelogo{Mechatronics}
\CopyrightLine{2014}{Published by Elsevier Ltd.}

\begin{document}
\begin{frontmatter}
\dochead{}
\title{Distributed nonlinear model predictive control of an autonomous tractor-trailer system}
\author[KULeuven]{Erkan Kayacan}
\author[NTU]{Erdal Kayacan}
\author[KULeuven]{Herman Ramon}
\author[KULeuven]{Wouter Saeys}
\address[KULeuven]{Department of Biosystems (BIOSYST), Division of Mechatronics, Biostatistics and Sensors (MeBioS), University of Leuven (KU Leuven), Kasteelpark Arenberg 30, 3001, Leuven, Belgium}
\address[NTU]{School of Mechanical \& Aerospace Engineering, Nanyang Technological University, 50 Nanyang Avenue, Singapore 639798}
\cortext[]{Corresponding author: Tel.: +32 16 3 77089; fax: +32 16 3 28590. \\ E-mail address: erkan.kayacan@biw.kuleuven.be (Erkan Kayacan)}

\begin{abstract}
This paper addresses the trajectory tracking problem of an autonomous tractor-trailer system by using a fast distributed nonlinear model predictive control algorithm in combination with nonlinear moving horizon estimation for the state and parameter estimation in which constraints on the inputs and the states can be incorporated. The proposed control algorithm is capable of driving the tractor-trailer system to any desired trajectory ensuring high control accuracy and robustness against environmental disturbances.
\end{abstract}
\begin{keyword}
agricultural robot, tractor-trailer system, autonomous vehicle, distributed nonlinear model predictive control, nonlinear moving horizon estimation.
\end{keyword}
\end{frontmatter}

\section{Introduction}\label{intro}
The basic idea behind automating agricultural production machines, \emph{e.g.} an autonomous tractor-trailer system, is not only the fact that energy and labour costs are increasing day by day but also farmers need durable accurate and reliable production machines. However, the steering accuracy of these machines decreases when the operator gets tired or has to perform other tasks apart from driving the tractor like operating mounted trailers. In such cases, advanced control algorithms are more than welcome. This has resulted in several automatic guidance systems, of which some are already available on the market.

Today's fast moving technology allows us the application of real time kinematic (RTK)-global positioning systems (GPSs) which can provide an accurate positioning accuracy of a few cm. Nonetheless the performance of the currently available machine guidance systems is rather limited due to the poor performance of the automatic control systems used for this purpose. The main reasons for this poor performance are the complex vehicle dynamics and the large variation in soil conditions which make that the conventional (\emph{e.g.} PID) controllers for machine guidance have to be tuned very conservatively. By conservative tuning, robustness of the controller is obtained at the price of performance. Moreover, the constraints of the mechanical system cannot be taken into account directly in these controllers, such that the ad hoc implementation of these constraints can lead to suboptimal behavior of the system. In such cases, advanced control algorithms which can deal with constraints on the states and the inputs are coherent preferences for the control of complex outdoor vehicles.

Applied to agricultural machinery, model predictive control (MPC) has several advantages over conventional controllers, e.g. they can deal with the constraints on the system and actuator saturation. The main goal of MPC is to minimize a performance criterion with respect to constraints of a system's inputs and outputs. The MPC caught the attention of researchers in the 1980s, and the first MPC controllers were implemented in the process industry which has less stringent real-time requirements due to large sampling periods in the order of seconds or minutes \cite{Jan}. The reason for a such a preference is that MPC depends upon repetitive online solution of an optimal control problem.

Large scale complex systems can be divided into a finite number of subsystems. Real-life applications may be continuous (power networks, sewer networks, water networks, canal and river networks for agriculture, etc) or discrete (traffic control, railway control, etc) \cite{Javalera}. The common approach to control these systems is the use of a decentralized control approach, \emph{e.g.} decentralized MPC in which the interactions between the subsystems are considered as disturbances to each subsystem. As these controllers are not aware of the interactions with other subsystems, they will exhibit selfish behavior leading to suboptimal performance of the global system. An alternative solution is the use of a centralized control approach, \emph{e.g.} centralized MPC. However, centralized MPC design for such a complex large scale system may not be practical due to the computational requirements or to the impossibility of obtaining a centralized model of the whole system including all the subsystem interactions. Besides, the computational complexity, another disadvantage of the centralized control approach is that all subsystems have to trust one central controller which is difficult to coordinate and maintain \cite{Stewart}. One way of addressing such problems is to use of distributed MPC in which the overall system is controlled by local MPCs based on a limited information about the system to be controlled or a partial state information \cite{Maestre}. Roughly speaking, whereas this type of control approach consisting of several local agents requires less computational power when compared to its centralized counterpart, the overall control accuracy of the system highly depends on the cooperation and communication between the local agents.

As an agricultural production machine, a tractor-trailer is a complex mechatronic system which consists of several subsystems that interact with each other as a result of energy flows. For instance,  the diesel engine, the steering system of the tractor and the steering system of the trailer share the same hydraulic oil. As a result, once an input is applied to one of the subsystems, it always affects the others. Considering the disadvantages of decentralized and centralized control approaches mentioned above, complex mechatronic systems, such as a tractor-trailer, are the worthwhile considering distributed control approach since the design of robust and accurate controllers for such systems is not a straightforward task due to their highly nonlinear dynamics \cite{Igreja}.

Researchers have recently been focussing on distributed control in which some limited information is transmitted among local agents. In distributed MPC, the optimization problem is broken into smaller pieces under the assumption of solving many small problems is faster and more scalable than solving one large problem \cite{Richards}.  A detailed survey  about the architectures for distributed and hierarchical MPC can be found in \cite{MPCservey}. There are two main approaches to distributed control: independent distributed control \cite{Dunbarcoupled,Dunbardecoupled} and cooperative distributed control \cite{Venkat,Rawlings}. While in the former, each subsystem agent considers network interactions only locally resulting into a Nash equilibrium for the performance of the system, in the latter, all local control actions actions are considered on all subsystems resulting into a Pareto optimum \cite{Ferramosca}. So, in independent distributed NMPC (iDiNMPC), the cost function of each subsystem consists of only the states of the local subsystem. On the other hand, in cooperative distributed NMPC (cDiNMPC), the cost function of each subsystem consist of the states of the overall system dynamics. An iterative cooperative distributed case was proposed in \cite{Stewart}. It has been shown in \cite{Stewart} that the communication between subsystems and using the global cost function result converging to the one of the corresponding centralized control case as the iterations number increases. Since the trajectory following accuracies of both the tractor and the trailer are essential in an agricultural operation, the latter approach is followed in this paper.


In this paper, a DiNMPC with the \emph{ACADO} code generation tool \cite{ACADO, Diehl} for the trajectory tracking problem of an autonomous tractor-trailer system has been developed and tested in real-time in the presence of several uncertainties, nonlinearities and biological variabilities. Although a tractor-trailer system is relatively less complex when compared to other large-scale systems (power networks, etc.), short times for optimization are crucial for such a mechatronic system. Since the optimization problem of NMPC is a complex problem and it is time-consuming, the main goal of this study is to design a fast NMPC for the tractor-trailer system. To succeed, the following selections have been made:
\begin{enumerate}
  \item The use of a kinematic model instead of a dynamic model
  \item The use of C++ source files to realize the control algorithm in real-time
  \item The use of the distributed control algorithm instead of a centralized one.
\end{enumerate}

Thanks to the selection above, the feedback times of the cDiNMPC and iDiNMPC are around 7ms and 3ms, respectively.

This paper has been organized as follows: The kinematic model of the system is presented in Section \ref{SystemModel}. The basics of the implemented DiNMPC and the learning process by using a nonlinear moving horizon estimation (NMHE) method have been explained in Section \ref{cDiNMPCmhe}.  The experimental set-up and the experimental results are described in Section \ref{realtime}. Finally, some conclusions have been drawn from this study in Section \ref{Conc}.

\section{Kinematic tricycle model of a tractor-trailer system }\label{SystemModel}
\begin{figure}[b!]
\centering
  \includegraphics[width=1.5in]{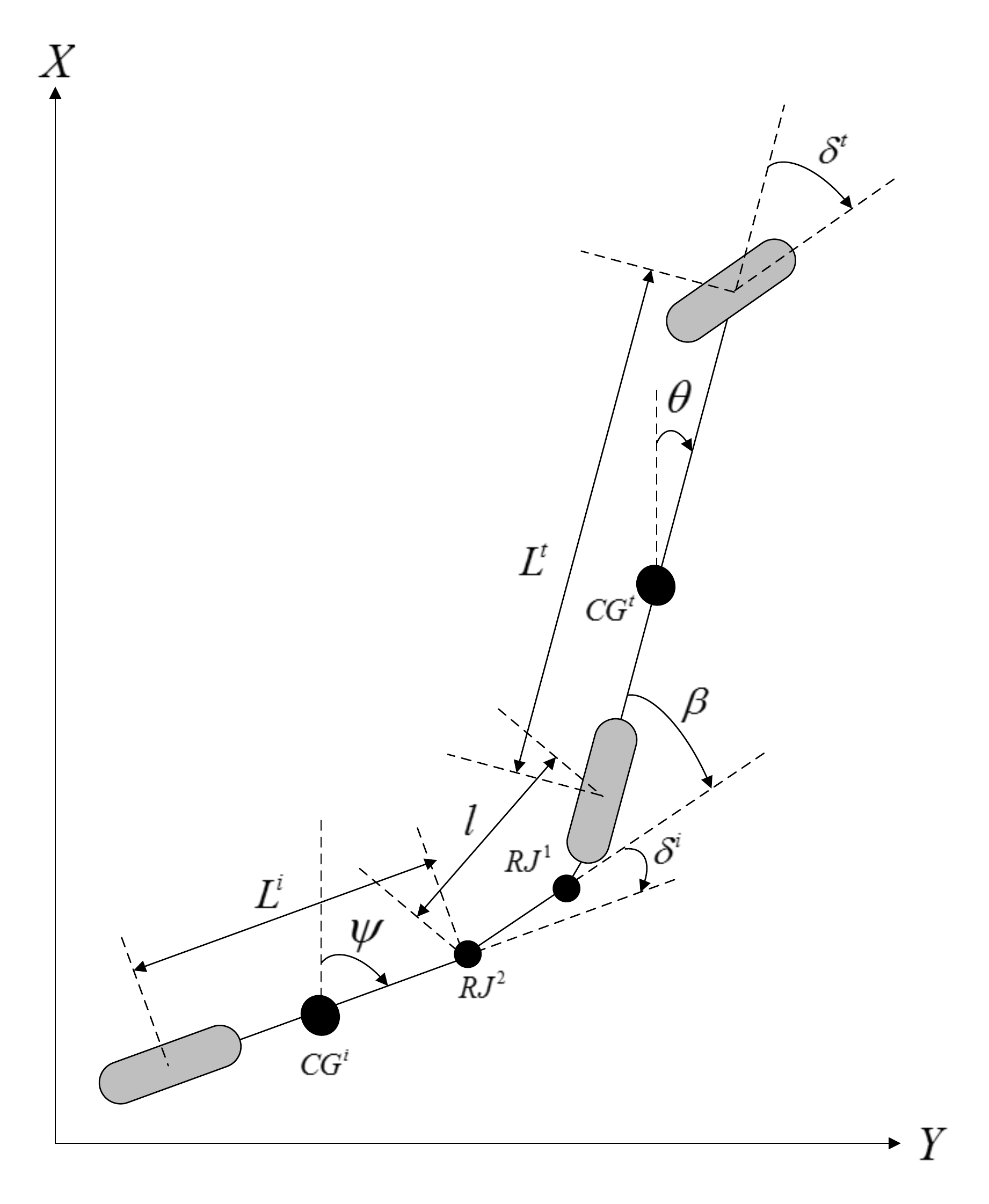}\\
  \caption{Schematic illustration of tricycle model for an autonomous tractor-trailer system}
  \label{kinematic}
\end{figure}

The schematic diagram of an autonomous tractor-trailer system is presented in Fig. \ref{kinematic}.

The model for the autonomous tractor-trailer system is \emph{a kinematic model} neglecting the dynamic force balances in the equations of motion. A dynamic model would, of course, represent the system behaviour with a better accuracy, but then a system identification and multibody modelling techniques would be needed for obtaining an accurate dynamic model of the system. Moreover, a dynamic model would increase the computational burden in the optimization process in DiNMPC. Thus, an extension of a simpler well-known tricycle kinematic model in \cite{karkeejournal,karkeephd} has been used for the DiNMPC design in this paper. The extensions are the additional three slip parameters ($\mu$, $\kappa$ and $\eta$) and the definition of the yaw angle difference between the tractor and the trailer by using two angle measurements ($\delta^{i}$ and $\beta$) instead of one angle measurement.

The equations of motion of the system to be controlled are as follows:
\small
\begin{eqnarray}\label{kinematicmodel}
\left[
  \begin{array}{c}
  \dot{x}^{t}  \\
  \dot{y}^{t}  \\
  \dot{\theta} \\
  \dot{x}^{i}  \\
  \dot{y}^{i}  \\
  \dot{\psi}   \\
  \end{array}
  \right]
=
\left[
  \begin{array}{c}
  \mu v \cos{(\theta)} \\
 \mu v \sin{(\theta)}  \\
\frac{\mu v \tan{ (\kappa \delta^{t}) } }{L^t} \\
 \mu v \cos{(\psi)}  \\
 \mu v \sin{(\psi)}  \\
\frac{\mu v }{L^i}\big(\sin{(\eta \delta^{i} + \beta)} - \frac{l }{L^t} \tan{ (\kappa \delta^{t})} \cos{(\eta \delta^{i} + \beta)} \big)  \\
  \end{array}
  \right]
\end{eqnarray}
\normalsize
where ${x}^{t}$ and $y^{t}$ represent the position of the tractor, $\theta$ is the yaw angle of the tractor, ${x}^{i}$ and $y^{i}$ represent the position of the trailer, $\psi$ is the yaw angle of the trailer, $v$ is the longitudinal speed of the system. Since the tractor and trailer rigid bodies are linked by two revolute joints at a hitch point, the tractor and the trailer longitudinal velocities are coupled to each other. The input to the tractor, the steering angle of the front wheel of the tractor, is represented by $\delta^{t}$, $\beta$ is the hitch point angle between the tractor and the drawbar at $RJ^{1}$; $\delta^{i}$ is the steering angle between the trailer and the drawbar at $RJ^{2}$ which is the input to the trailer; $\mu$, $\kappa$ and $\eta$ are the slip coefficients for the wheel slip of the tractor, side-slip for the tractor and side-slip for the trailer, respectively.

The physical parameters that can be directly measured are as follows: $L^t=1.4m$, $L^i=1.3$, and $l=1.1$.  The parameters $L^t$, $L^i$, and $l$ represent the distance between the front axle of the tractor and the rear axle of the tractor, the distance between $RJ^2$ and the rear axle of the trailer, and the distance between $RJ^1$ and $RJ^2$, respectively.

\section{Distributed nonlinear model predictive control and nonlinear moving horizon estimation framework} \label{cDiNMPCmhe}
\subsection{Distributed nonlinear model predictive control}
In order to be able to design a DiNMPC, partitioned model of large-scale systems should be available derived from partitioning methods as non-overlapping decomposition or completely overlapping decomposition. However, considering the kinematic model in \eqref{kinematicmodel}, the first three states of \eqref{kinematicmodel} are the state equations of the tractor while the last three states are the state equations of the trailer. Thus, the equations of motion for the tractor-trailer system represented in \eqref{kinematicmodel} are naturally decoupled so that partitioning methods are not needed for our system. In addition, distributed solutions can be proposed for such decoupled models.

In NMPC, it is to be noted that the computational load in the optimization process increases when the number of the states of the system increases. For this reason, one of the advantages of the DiNMPC is to lower the computational burden by decoupling the subsystems. Once the subsystems are decoupled from each other, the DiNMPC algorithm will not need a scalable memory load. A second advantage is that less communication between subsystems results in the reduction of the transmission delays and overloads.

For the formulation of DiNMPC, it is assumed that the plant comprises only two subsystems due to the fact that the tractor-trailer system has two subsystems.
\subsubsection{Model}
A totally decoupled nonlinear model can be written in the following form:
\begin{equation}
\dot{x}=f(x,u)
\end{equation}
where
 \begin{eqnarray*}\label{}
x=\left[
  \begin{array}{c}
  x_1  \\
  x_2  \\
  \end{array}
  \right], \;
u=\left[
  \begin{array}{c}
  u_1  \\
  u_2  \\
  \end{array}
  \right], \;
  f(x,u)=\left[
  \begin{array}{c}
  f_1(x_1,u_1,u_2)  \\
  f_2(x_2,u_1,u_2)  \\
  \end{array}
  \right]
  \end{eqnarray*}
for which $x$ $\in$ $\mathbb{R}^n$, $u$ $\in$ $\mathbb{R}^m$, and $f:\mathbb{R}^n \times \mathbb{R}^m \longrightarrow \mathbb{R}^n$.
\subsubsection{Constraints}
At each time step, the inputs have to satisfy:
\begin{equation}
u_1 \in \mathbb{U}_1, \;\;\; u_2 \in \mathbb{U}_2
\end{equation}
where $\mathbb{U}_1$ and $\mathbb{U}_2$ are compact and convex. The constraints are defined uncoupled because the feasible regions of the inputs do not affect each other.

\subsubsection{Objective functions}
The stage cost and the terminal penalty are respectively written for each subsystem $i \in \mathbb{I}_{1:2}$ as follows:
\begin{eqnarray}
J_{iSC} (x_i, u_i) & = & \| x_{ir} (t) - x_i (t) \|^{2}_{Q_i} + \| u_{ir} (t) - u_i (t) \|^{2}_{R_i}  \\
J_{iTP} (x_i)      & = & \| x_{ir} (t_k+t_h) - x_i (t_k+t_h) \|^{2}_{S_i}
\end{eqnarray}
where $Q_i \in \mathbb{R}^{n_i \times n_i}$, $R_i \in \mathbb{R}^{m_i \times m_i}$ and $S_i \in \mathbb{R}^{n_i \times n_i}$ are weighting matrices being symmetric and positive definite, $x_{ir}$ and $u_{ir}$ are the references for the states and the inputs, $x_i$ and $u_i$ are the states and the inputs, $t_k$ stands for the current time, $t_h$ is the prediction horizon.

The objective function for each subsystem $i \in \mathbb{I}_{1:2}$ is written as follows:
\begin{equation}
 \begin{aligned}
 & J_i (x_i, u_1, u_2)=
 & & \int^{t_k+t_h}_{t_k} \big(J_{iSC} (x_i, u_i)\big) dt + J_{iTP} (x_i)  \\
 & && \forall t \in [t_k, t_k+t_h] \\
  \end{aligned}
\end{equation}

\subsubsection{Formulation for cooperative DiNMPC}
For the cDiNMPC case, it is to be noted that the states $x_{i}$ are the functions of the inputs $u_1$ and $u_2$. Thus, the objective function $J_i$ becomes a function of the inputs $u_1$ and $u_2$. Afterwards, the plant objective function can be written as follows:
\begin{equation}
J (x_1, x_2, u_1, u_2)= \rho_{1} J_1 (x_1, u_1, u_2) + \rho_{2} J_2 (x_2, u_1, u_2)
\end{equation}
where $\rho_1, \rho_2 > 0 $ are the weighting coefficients.

The cDiNMPC formulation for the first subsystem is as follows:
\begin{equation}
 \begin{aligned}
 & \underset{x_1(.), x_2(.), u_1(.), u^{*}_2(.)}{\text{min}}
 & & J(x_1,x_2,u_1,u^{*}_2) \\
 & \text{subject to}
 && x_1(t_k) = \hat{x}_1(t_k) \\
 && & x_2(t_k) = \hat{x}_2(t_k) \\
 && & \dot{x}_1(t) = f_1 \big(x_1(t), u_1(t), u^{*}_2(t) \big) \\
 && & \dot{x}_2(t) = f_2 \big(x_2(t), u_1(t), u^{*}_2(t) \big) \\
 && & x_{1_{min}} \leq x_1(t) \leq x_{1_{max}} \\
 && & x_{2_{min}} \leq x_2(t) \leq x_{2_{max}} \\
 && & u_{1_{min}} \leq u_1(t) \leq u_{1_{max}} \\
  \end{aligned}
  \label{cDiNMPC1}
\end{equation}
where  $J$ is the plant objective function, $x_1$ and $x_2$ are the states, and $u_1$ and $u_2$ are the inputs, Moreover, upper and lower bounds on the states and the input are represented by $x_{1_{min}}$, $x_{1_{max}}$, $x_{2_{min}}$, $x_{2_{max}}$ $u_{1_{min}}$ and $u_{1_{max}}$. The input of the second subsystem $u^{*}_2$ represents the prediction values which are coming from the cDiNMPC for second subsystem.

Similarly, the cDiNMPC formulation for the second subsystem is as follows:
\begin{equation}
 \begin{aligned}
 & \underset{x_1(.), x_2(.), u^{*}_1(.), u_2(.)}{\text{min}}
 & & J(x_1,x_2,u^{*}_1,u_2) \\
 & \text{subject to}
 && x_1(t_k) = \hat{x}_1(t_k) \\
 && & x_2(t_k) = \hat{x}_2(t_k) \\
 && & \dot{x}_1(t) = f_1 \big(x_1(t), u^{*}_1(t), u_2(t) \big) \\
 && & \dot{x}_2(t) = f_2 \big(x_2(t), u^{*}_1(t), u_2(t) \big) \\
 && & x_{1_{min}} \leq x_1(t) \leq x_{1_{max}} \\
 && & x_{2_{min}} \leq x_2(t) \leq x_{2_{max}} \\
 && & u_{2_{min}} \leq u_2(t) \leq u_{2_{max}} \\
  \end{aligned}
  \label{cDiNMPC2}
\end{equation}
where the upper and lower bounds on the input of the second system are represented by $u_{2_{min}}$ and $u_{2_{max}}$. The input of the first subsystem $u^{*}_1$ represents the prediction values which are coming from the cDiNMPC for first subsystem.

\emph{Remark:} If the equations of motion for the tractor-trailer system represented in \eqref{kinematicmodel} are carefully considered, it can be easily seen that the input to the second subsystem (the trailer) does not appear in the equations of motion of the first subsystem (the tractor). As a result, the objective function of the first subsystem cannot be minimized with respect to the input to the second subsystem. Thus, \eqref{cDiNMPC2} is not valid for our real-time system. This fact leads us to design an iDiNMPC, which is formulated in Section \ref{iDiNMPCsection}, for the trailer.

\subsubsection{Formulation for independent DiNMPC}
\label{iDiNMPCsection}

For the iDiNMPC case, the plant objective function consists of only $J_2 (x_2, u_1, u_2)$. The iDiNMPC formulation for the trailer is as follows:
\begin{equation}
 \begin{aligned}
 & \underset{x_2(.), u^{*}_1(.), u_2(.)}{\text{min}}
 & & J_{2}(x_2,u^{*}_1,u_2) \\
 & \text{subject to}
 && x_2(t_k) = \hat{x}_2(t_k) \\
 && & \dot{x}_2(t) = f_2 \big(x_2(t), u^{*}_1(t), u_2(t) \big) \\
 && & x_{2_{min}} \leq x_2(t) \leq x_{2_{max}} \\
 && & u_{2_{min}} \leq u_2(t) \leq u_{2_{max}} \\
  \end{aligned}
  \label{iDiNMPC}
\end{equation}

As can be seen from \eqref{iDiNMPC}, the objective function consists of only the states of the second subsystem (the trailer) and the inputs of the first and second subsystems. Since there are no states of the first subsystem (the tractor), this objective function can be minimized with respect to the second input to the second subsystem.

\subsection{Nonlinear moving horizon estimation}
Unlike the well known estimation method namely extended Kalman filter (EKF), NMHE treats the state and the parameter estimation within the same problem and also constraints can be incorporated \cite{TomKraus}. The constraints play an important role in the autonomous tractor-trailer system. For instance, the slip coefficients in  \eqref{constraints2} cannot be bigger than $1$.

The NMHE problem can be formulated as follows:
\begin{equation}
 \begin{aligned}
 & \underset{x(.),p,u(.)}{\text{min}}
 & & \int^{t_k}_{t_k-t_h} \big(\| y_m (t) - y (t) \|^{2}_{V_y} + \| u_m (t) - u (t) \|^{2}_{V_u} \big) dt  \\ &
 && + \left\|
  \begin{array}{c}
    \hat{x} (t_k-t_h) -x (t_k-t_h)  \\
    \hat{p} - p
 \end{array}
 \right\| ^{2}_{V_s} \\
 & \text{subject to}
 &&  \dot{x}(t) = f \big(x(t),u(t),p \big) \\
 &&& x_{min} \leq x(t) \leq x_{max} \\
 &&& p_{min} \leq p \leq p_{max} \;\; \text{for all} \;\; t \in [t_k - t_h ,t_k]\\
  \end{aligned}
    \label{mhe}
\end{equation}
where $y_m$ and $u_m$ are the measured outputs and inputs, respectively. Deviations of the first states in the moving horizon window and the parameters from priori estimates $\hat{x}$ and $\hat{p}$ are penalised by a symmetric positive definite matrix $V_s$. Moreover,  deviations of the predicted system outputs and the measured outputs and deviations of the predicted system inputs and the measured inputs are penalised by symmetric positive definite matrices $V_y$ and $V_u$, respectively \cite{Ferreau}. Upper and lower bounds on the model parameters are represented by parameters $p_{min}$ and $p_{max}$, respectively.

The last term in the objective function in \eqref{mhe} is called the arrival cost. The reference estimated values $\hat{x} (t_k-t_h)$ and $\hat{p}$ are taken from the solution of NMHE at the previous estimation instant. In this paper, the arrival cost matrix $V_s$ has been chosen as a so-called smoothed EKF-update based on sensitivity information obtained while solving the previous NMHE problem \cite{Robertson}.

\subsection{Solution methods}

The optimization problems in DiNMPC \eqref{cDiNMPC1} and \eqref{iDiNMPC}, and in NMHE \eqref{mhe} are similar to each other, which makes that the same solution method can be applied for both DiNMPC and NMHE problems \cite{TomKraus}. In this paper, the multiple shooting method has been used in a fusion with a generalized Gauss-Newton method. Although the number of iterations cannot be determined in advance, a simple solution was proposed in \cite{Diehl2} in which the number of Gauss-Newton iterations is limited to $1$. Meanwhile, each optimization problem is initialized with the output of the previous one.

The \emph{ACADO} code generation tool, an open source software package for optimization problems \cite{ACADO,Houska2011a}, has been used to solve the constrained nonlinear optimization problems in the NMPC and NMHE. First, this software generates C-code, then the auto-generated code is converted into a \emph{.dll} file to be used in \emph{LabVIEW}. Detailed information about the \emph{ACADO} code generation tool can be found in \cite{home}.

\section{Experimental set-up description and real-time results} \label{realtime}
\subsection{Experimental set-up description}
\begin{figure}[b!]
\centering
  \includegraphics[width=3in]{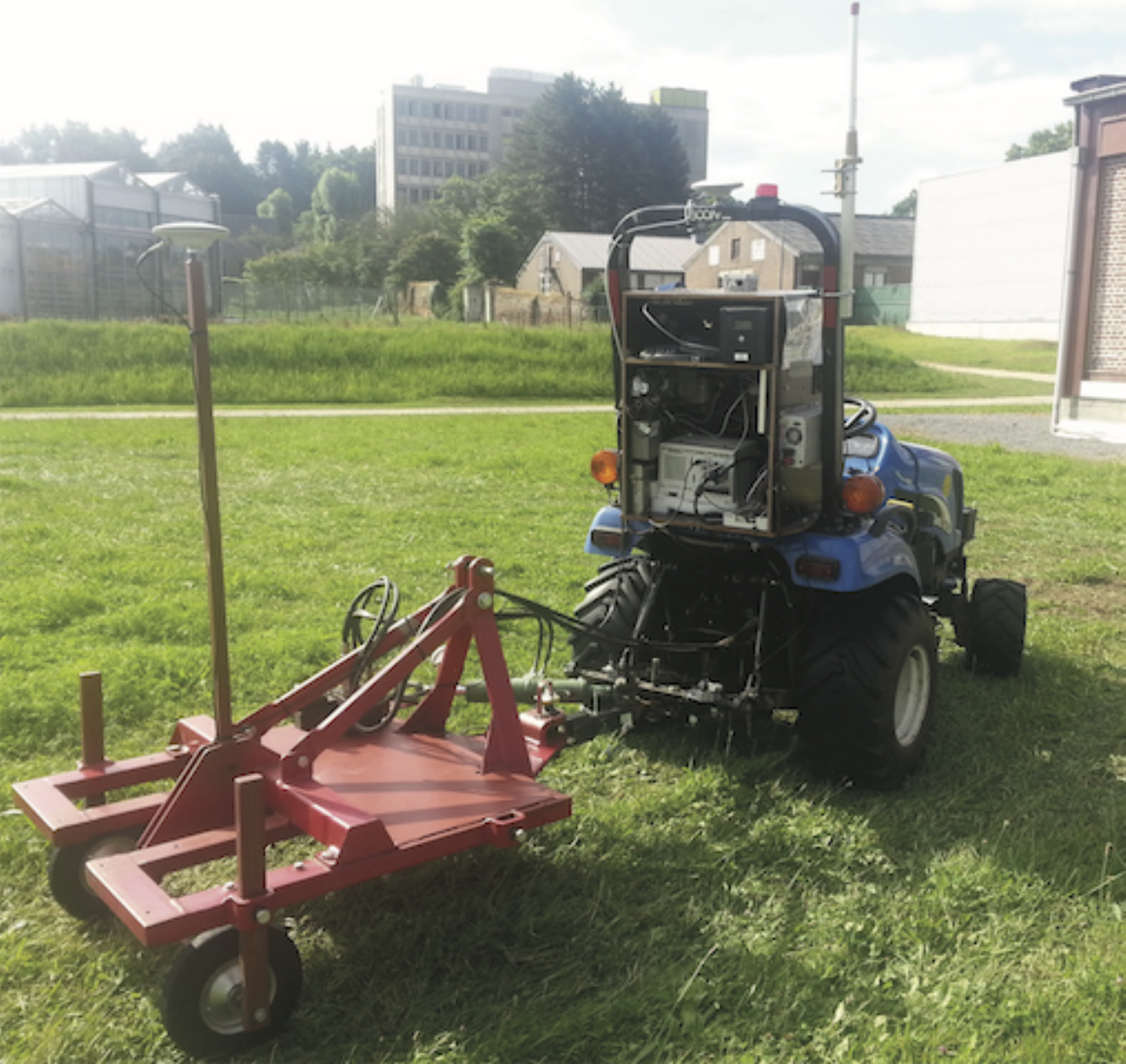}\\
  \caption{The tractor-trailer system}
  \label{tractor1}
\end{figure}

The global aim of the real-time experiments in this paper is to track a space-based trajectory with a small agricultural tractor-trailer system shown in Fig. \ref{tractor1}. Two GPS antennas are located straight up the center of the tractor rear axle and the center of the trailer to provide highly accurate positional information. They are connected to a Septentrio AsteRx2eH RTK-DGPS receiver (Septentrio Satellite Navigation NV, Belgium) with a specified position accuracy of 2cm at a 20-Hz sampling frequency. The Flepos network supplies the RTK correction signals via internet by using a \emph{Digi Connect WAN 3G} modem.

The GPS receiver and the internet modem are connected to a real time operating system (PXI platform, National Instrument Corporation, USA) through an RS232 serial communication. The PXI system acquires the steering angles and the GPS data, and controls the tractor-trailer system by sending messages to actuators. A laptop connected to the PXI system by WiFi functions as the user interface of the autonomous tractor. The control algorithms are implemented in $LabVIEW^{TM}$ version 2011, National Instrument, USA. They are executed in real time on the PXI and updated at a rate of 5-Hz.

The cDiNMPC and the iDiNMPC calculate the desired steering angles for the front wheels of the tractor and the actuator of the trailer respectively, and two low level controllers, PI controllers in our case, are used to control the steering mechanisms. While the position of the front wheels of the tractor is measured using a potentiometer mounted on the front axle yielding a position measurement resolution of $1^{\circ}$, the position of the electro-hydraulic valve on the trailer is measured by using an inductive sensor with $1^{\circ}$ precision.

\begin{figure}[h!]
\centering
\subfigure[ ]{
\includegraphics[width=2.5in]{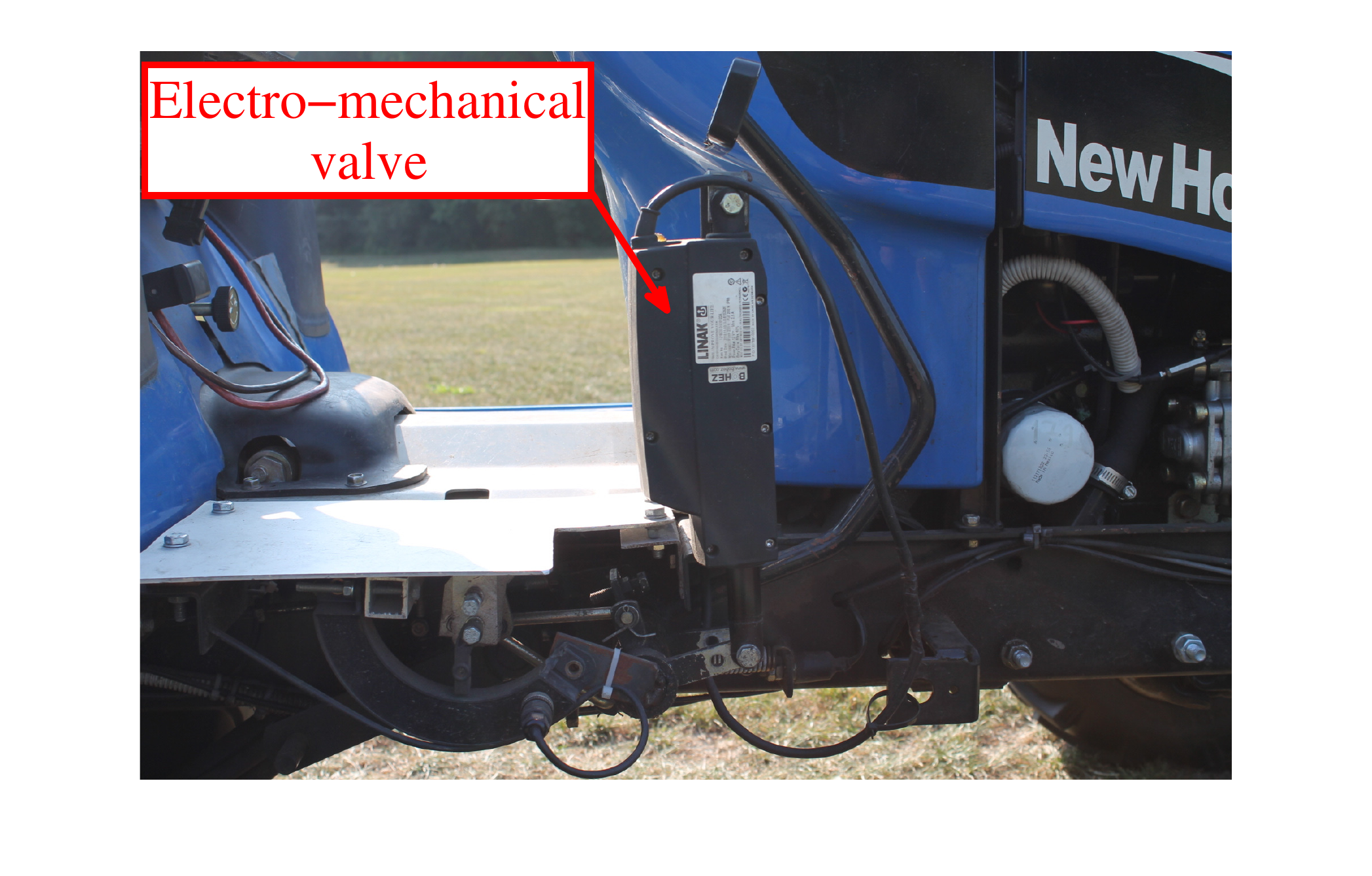}
\label{hydrostate_EMV}
}
\subfigure[ ]{
\includegraphics[width=2.5in]{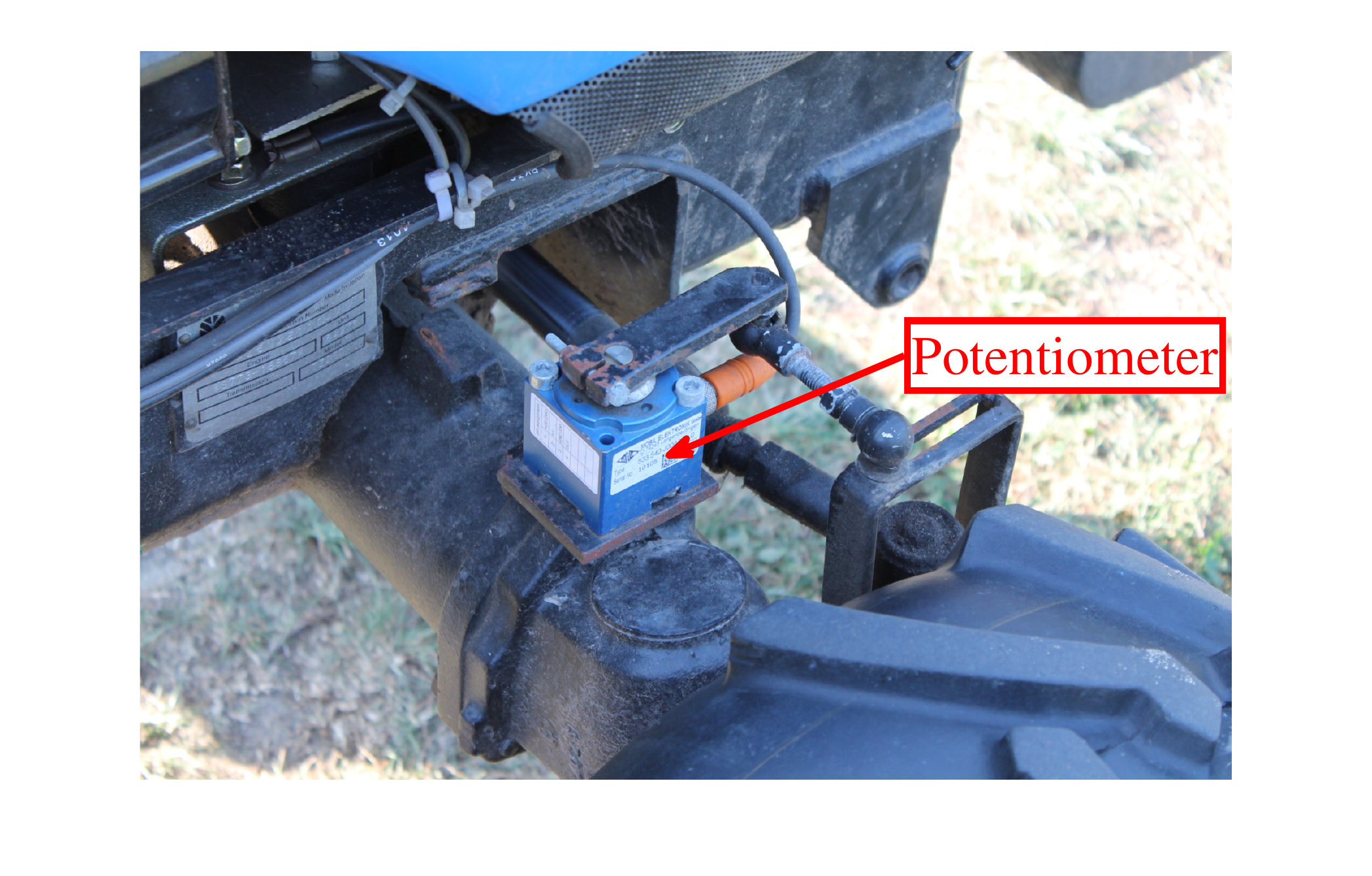}
\label{steering-potantiometer}
}
\subfigure[ ]{
\includegraphics[width=2.5in]{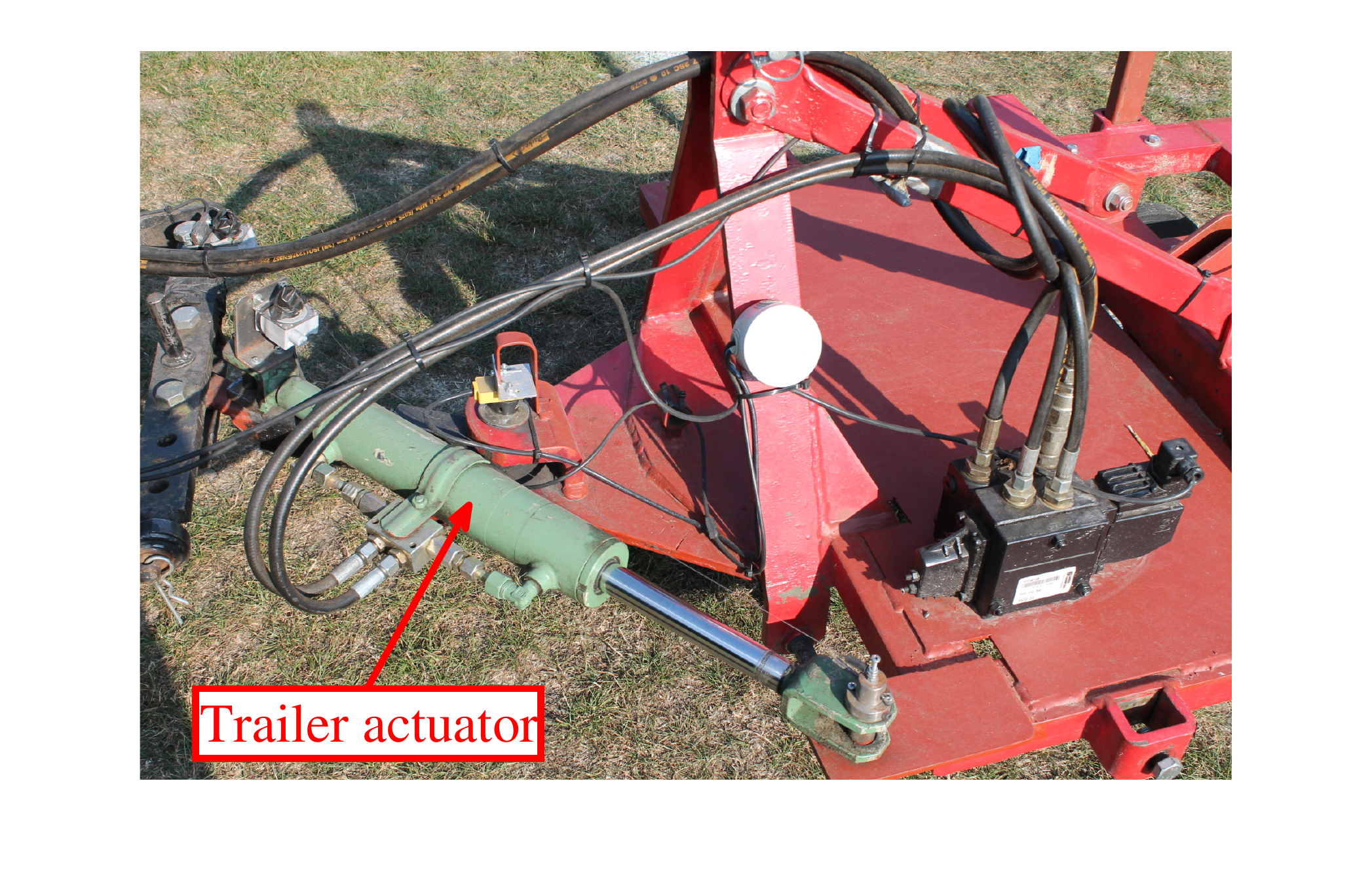}
\label{Trailer actuator}
}
\caption[Optional caption for list of figures]{Illustration of the actuators which have been added to the small-scale tractor to make it autonomous: Hydrostat electro-mechanical valve (a) Steering angle potentiometer (b) Trailer actuator (c)}
\label{sensors}
\end{figure}

The speed of the tractor is controlled by using an electro-mechanic valve. A cascade system with two PID controllers are built in the speed control system. The PID controllers in outer closed-loop and inner closed-loop are generating the desired pedal position with respect to the speed of the tractor and the voltage for the electro-mechanic valve for the pedal position, respectively. The hydrostat electro-mechanical valve (Fig. \ref{hydrostate_EMV}), the steering angle potentiometer (Fig. \ref{steering-potantiometer}) and the trailer actuator (Fig. \ref{Trailer actuator}) are shown in Fig. \ref{sensors} respectively.

\subsection{Implementation of NMHE}
Some states of the autonomous tractor-trailer system cannot be measured. Even if states can be measured directly, the obtained measurements contain  time delays and are contaminated with noise. Sometimes, data loss from the GPS for global localization of the tractor occur. In order to estimate the unmeasurable states or parameters, the NMHE method is used. Since only one GPS antenna is mounted on the tractor and one GPS antenna on the trailer, the yaw angles of the tractor and the trailer cannot be measured. As knowledge of the yaw angles of tractor and trailer is essential for accurate trajectory tracking, these have to be accurately estimated.

The inputs to the NMHE algorithm are the position of the tractor, the longitudinal velocity values from the encoders mounted on the rear wheels of the tractor and the steering angle values from the potentiometer mounted on the steering axle of the front wheels of the tractor, the position of the trailer and the steering angle values from the inductive sensors on the trailer and the hitch point angle from the potentiometer between the tractor and the drawbar. The outputs of NMHE are the positions of the tractor and the trailer in the x- and y-coordinate system, the yaw angles for both the tractor and the trailer, the slip coefficients, the hitch point angle and the longitudinal speed. In all the real-time experiments, the estimated values are fed to the DiNMPC.

The NMHE problem is solved at each sampling time with the following constraints on the parameters:
\begin{eqnarray}\label{constraints2}
0.25  \leq & \mu & \leq 1 \nonumber \\
0.25  \leq & \eta & \leq 1 \nonumber \\
0.25  \leq & \kappa & \leq 1
\end{eqnarray}

The standard  deviations of the measurements are set to $\sigma_{x^t} = \sigma_{y^t} = \sigma_{x^i} = \sigma_{y^i} = 0.03$ m, $\sigma_{\beta} = 0.0175$ rad, $\sigma_{v} = 0.1$ m/s, $\sigma_{\delta^{t}} = 0.0175$ rad and $\sigma_{\delta^{i}} = 0.0175$ rad based on the information obtained from the real-time experiments. Thus, the following weighting matrices $V_y$ and $V_u$ have been used in NMHE:
\begin{eqnarray}\label{weightingmatricesVyVu}
V_{y} & = & diag(0.03,0.03,0.03,0.03,0.0175,0.01)^T \nonumber \\
V_{u} & = & diag(0.0175,0.0175)^T
\end{eqnarray}

\subsection{Implementation of DiNMPC}

The DiNMPC problems for the two subsystem are solved at each sampling time with the following constraints on the inputs, which are the steering angles of the tractor and the trailer:
\begin{eqnarray}\label{constraints}
-35 \degree  \leq & \delta^{t}(t) & \leq 35 \degree \nonumber \\
-25 \degree  \leq & \delta^{i}(t) & \leq  25 \degree
\end{eqnarray}

The references for the positions and the inputs of the tractor and trailer are respectively changed online while all other references are set to zero as follows:
\begin{eqnarray}\label{}
y_{1ref} & = & (x^t_{ref},y^t_{ref},0)^T \nonumber \\
u_{1ref} & = & (\delta^{t}_{ref}) \nonumber \\
y_{2ref} & = & (x^i_{ref},y^i_{ref},0)^T \nonumber \\
u_{2ref} & = & (\delta^{i}_{ref})
\end{eqnarray}

The input references are the recent measured the steering angle of the front wheel of the tractor and the steering angle of the trailer. They are used in the objective function to provide a possibility to penalize the variations of the inputs from timestep to timestep. Moreover, the weighting matrices $Q_i$, $R_i$ and $S_i$ are defined as follows:
\begin{eqnarray}\label{weightingmatricesQRS}
Q_i & = & diag(0.5,0.5,0) \nonumber \\
R_i & = & 5 \nonumber \\
S_i & = & diag(5,5,0)
\end{eqnarray}

As can be seen from \eqref{weightingmatricesQRS}, the weighting for the inputs has been chosen big enough in order to get well damped closed-loop behaviour. The reason for such a selection is that since the tractor-trailer system is slow, it cannot give a fast response. Moreover, the weighting matrix $S_i$ is set $10$ times bigger than the weighting matrix $Q_i$. Thus, the deviations of the predicted values at the end of the horizon from its reference are penalized $10$ times more in the DiNMPC cost function than the previous points.

\subsection{Real-time results}

A space-based trajectory consisting of three straight lines and two smooth curves has been used as a reference signal. Since the radius of the curves is equal to $10$m, the curvature of the smooth curves is equal to $0.1$ (The curvature of a circle is the inverse of its radius).

The reference generation method in this paper is as follows: As soon as the tractor starts off-track, first, it quickly calculates the closest point on the space-based trajectory, then it determines its desired point. The desired point is a fixed forward distance from the closest point on the trajectory at every specific time instant. While the selection of a big distance from the closest point on the trajectory  results in a steady-state error on the trajectory to follow while the drawback of selecting a small distance is that it results in oscillatory behavior of the steering mechanism. The main goal of the reference generation algorithm for the tractor is both preventing the oscillations of the steering mechanism and minimizing the steady-state trajectory following error. The optimum distance, which has been calculated by trial-and-error method in this paper, is $1.6$m ahead of the front axle of the tractor.

\begin{figure}[b!]
\centering
\includegraphics[width=3.5in]{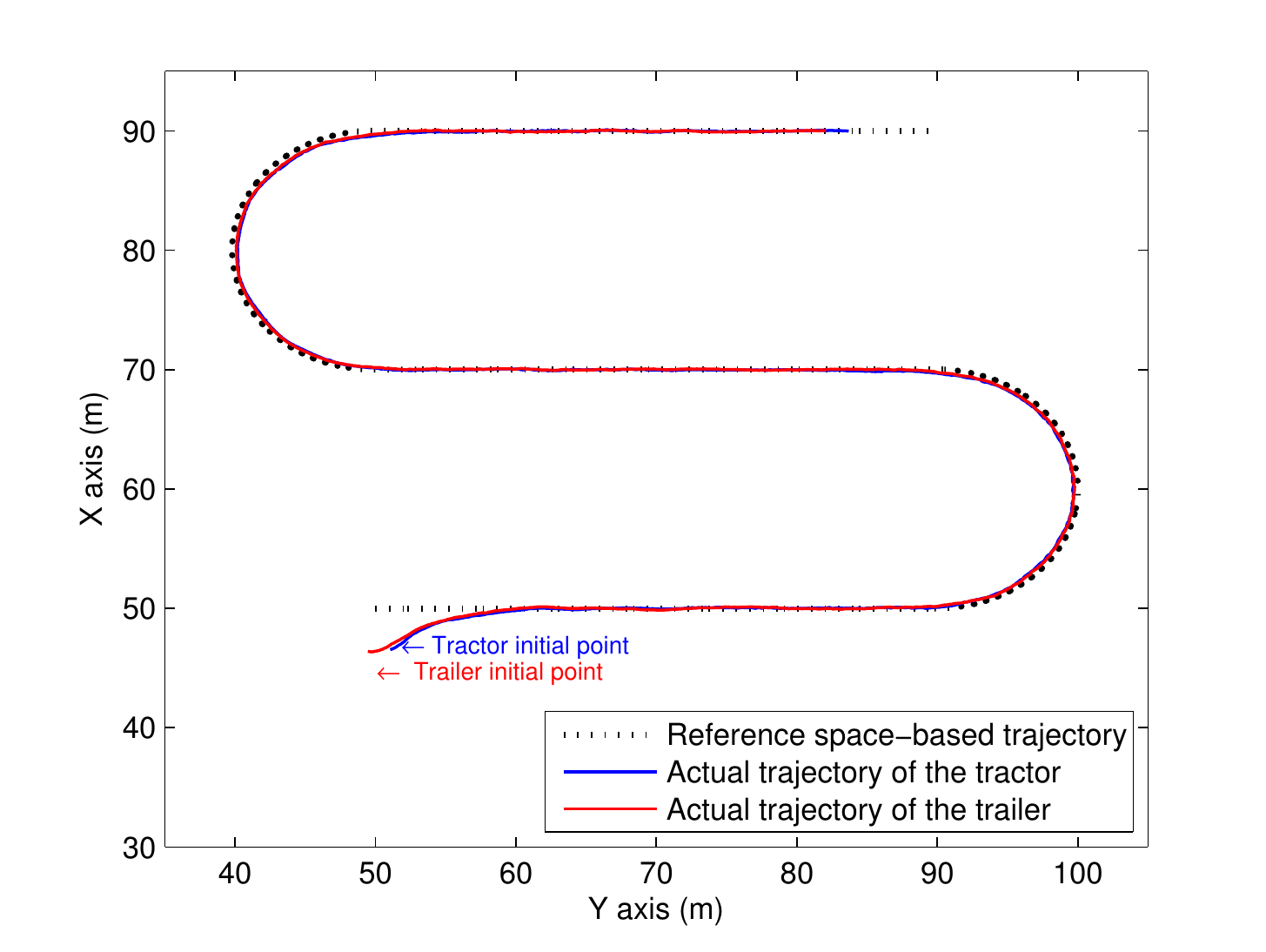}\\
\caption{Reference trajectory, vehicle actual trajectory}
\label{traj}
\end{figure}

As can be seen from Fig. \ref{traj}, the autonomous tractor-trailer system is capable to stay on-track after a finite time. In theory, since the reference generation algorithm places the target point $1.6$ meters ahead from the front axle of the tractor, there will be always a steady-state error for the curvilinear trajectories. On the other hand, no steady-state error is expected for the linear trajectories.

In the cDiNMPC case for the tractor, the weighting matrices are set to the same values to evaluate the effect of the weighting coefficients $\rho_1$ and $\rho_2$. During the real-time experiments, it is observed that the tractor shows oscillatory behaviour when the weighting coefficient $\rho_2$ for the objective function of the trailer is bigger than or equal to $\rho_1$. The reason is that since the first cDiNMPC tries to minimize the error of the tractor and the error of the trailer simultaneously, it is unable to generate a proper reference steering angle for the tractor. When the weighting coefficient $\rho_2$ is set to be smaller than $\rho_1$, the cDiNMPC is able to generate a proper steering angle for the tractor. In real-time experiments, the weighting coefficients $\rho_1$ and $\rho_2$ are set to $0.9$ and $0.1$, respectively. If a faster actuator for the trailer was available, more balanced selection of $\rho_1$ and $\rho_2$ could have been set. The control horizon $t_h$ is set to $3$ seconds which was found by trial-and-error method.

\begin{figure}[h!]
\centering
\includegraphics[width=3.5in]{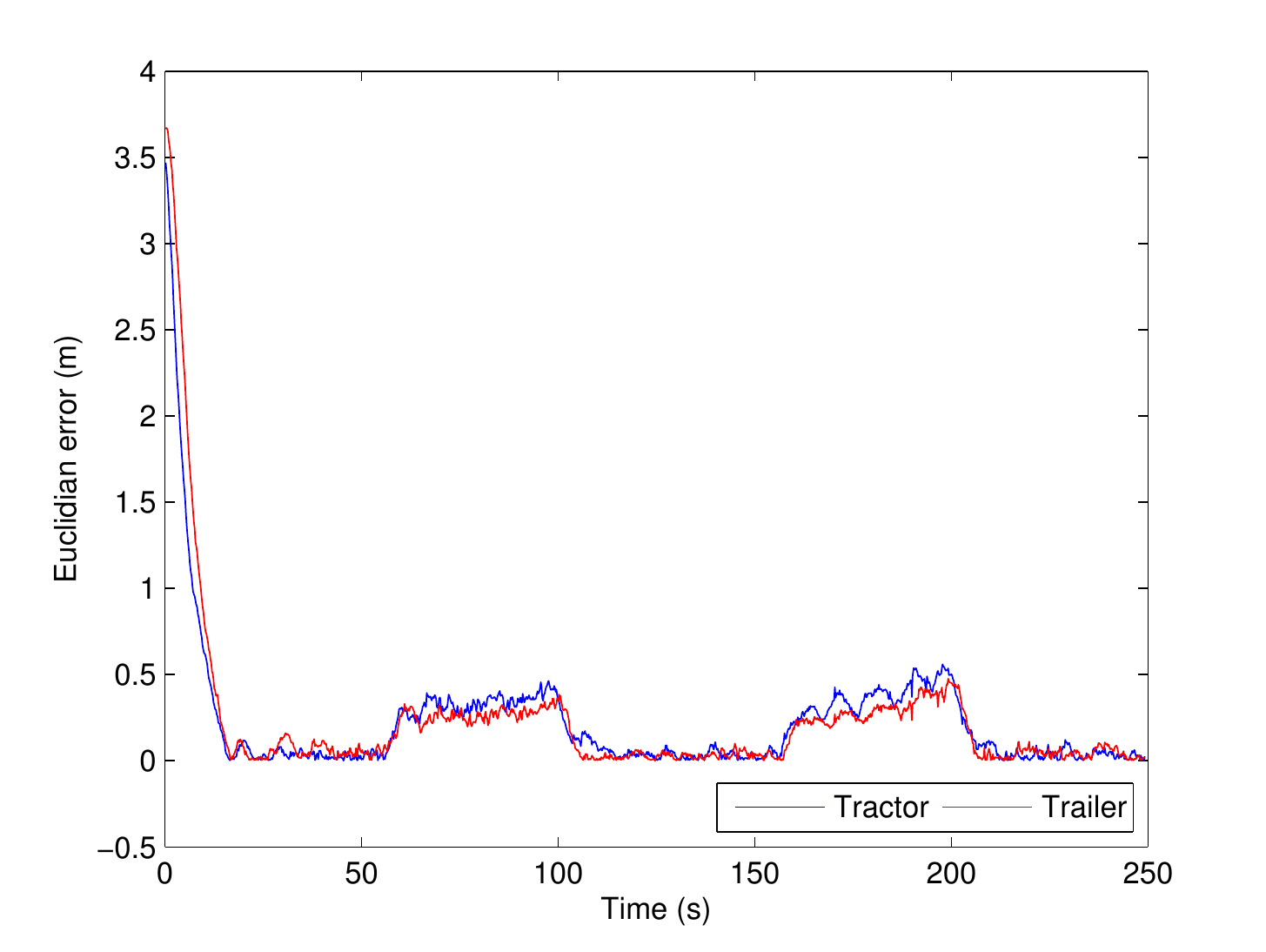}\\
\caption{Euclidian error to the space-based reference trajectory}
\label{error}
\end{figure}

Figure \ref{error} shows the Euclidian error to the space-based reference trajectory for both the tractor and the trailer. The mean values of the Euclidian error of the tractor and the trailer for the straight lines are $3.33$ cm and $3.22$ cm, respectively. Besides, the mean values of the Euclidian error of the tractor and the trailer for the curved lines are $36.20$ cm and $28.65$ cm, respectively. Although the DiNMPC calculates the proper outputs for $\delta^{i}$ at $RJ^2$, the error correction for the trailer is limited due to the fact that the length of the drawbar between the tractor and the trailer is only $20$ cm, which corresponded to a maximal lateral displacement of the trailer with respect to the tractor of $10.5$ cm.

\begin{figure}[h!]
\centering
\includegraphics[width=3.5in]{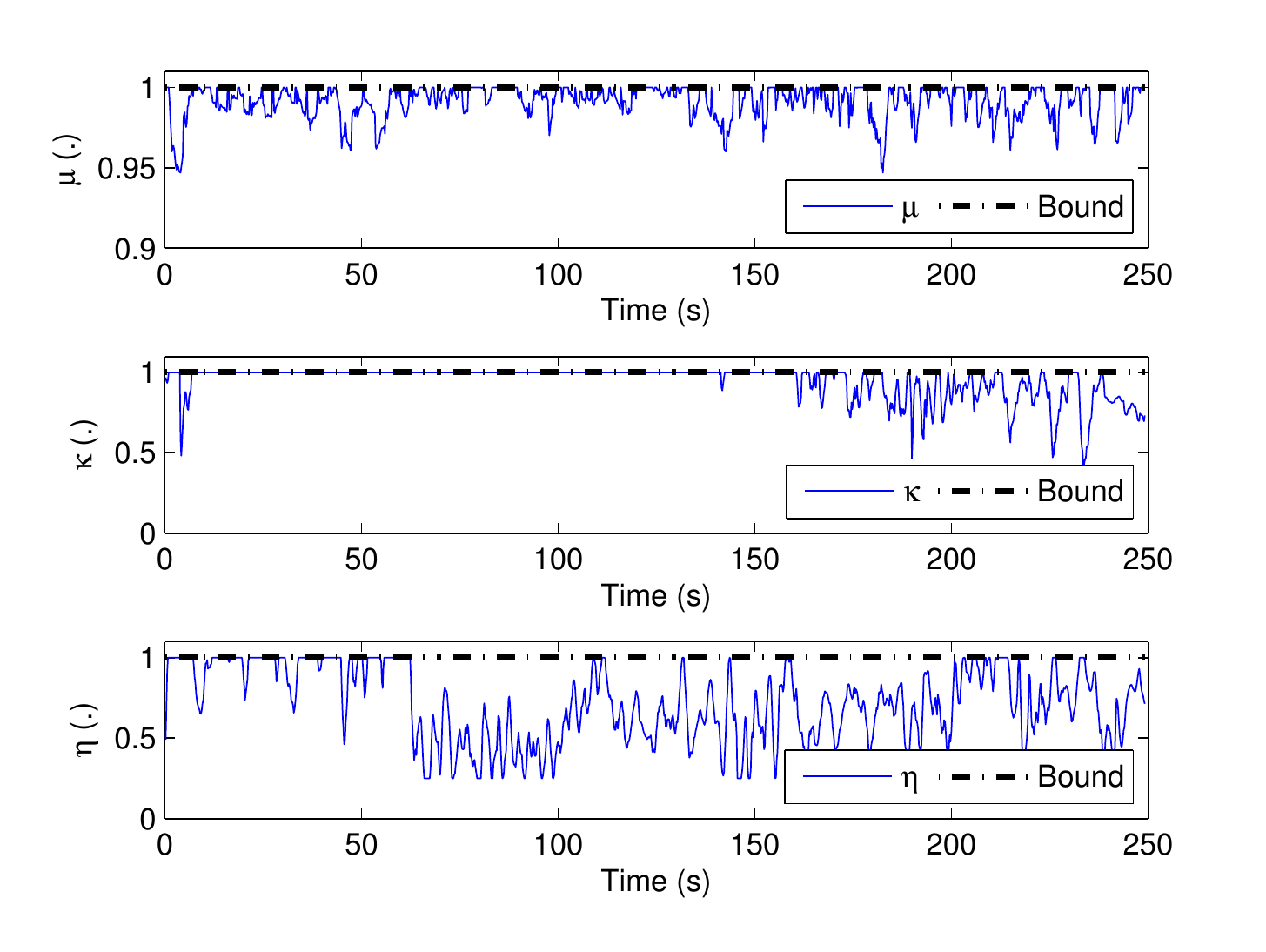}\\
\caption{Tractor longitudinal slip coefficient $(\mu)$, tractor ($\kappa$) and trailer ($\eta$) side slip coefficients}
\label{slips}
\end{figure}

Figure \ref{slips} represents the NMHE parameter estimation performance for the slip coefficients. As can be seen from this figure, the estimated parameter values are within the constraints in specified \eqref{constraints2}. Deviations in slip parameters occur when a vehicle accelerates, decelerates or soil conditions change, etc. However, this is not the case in our system. Instead, the deviations in the slip parameters are momentous due to modeling errors in our case.

\begin{figure}[h!]
\centering
\includegraphics[width=3.5in]{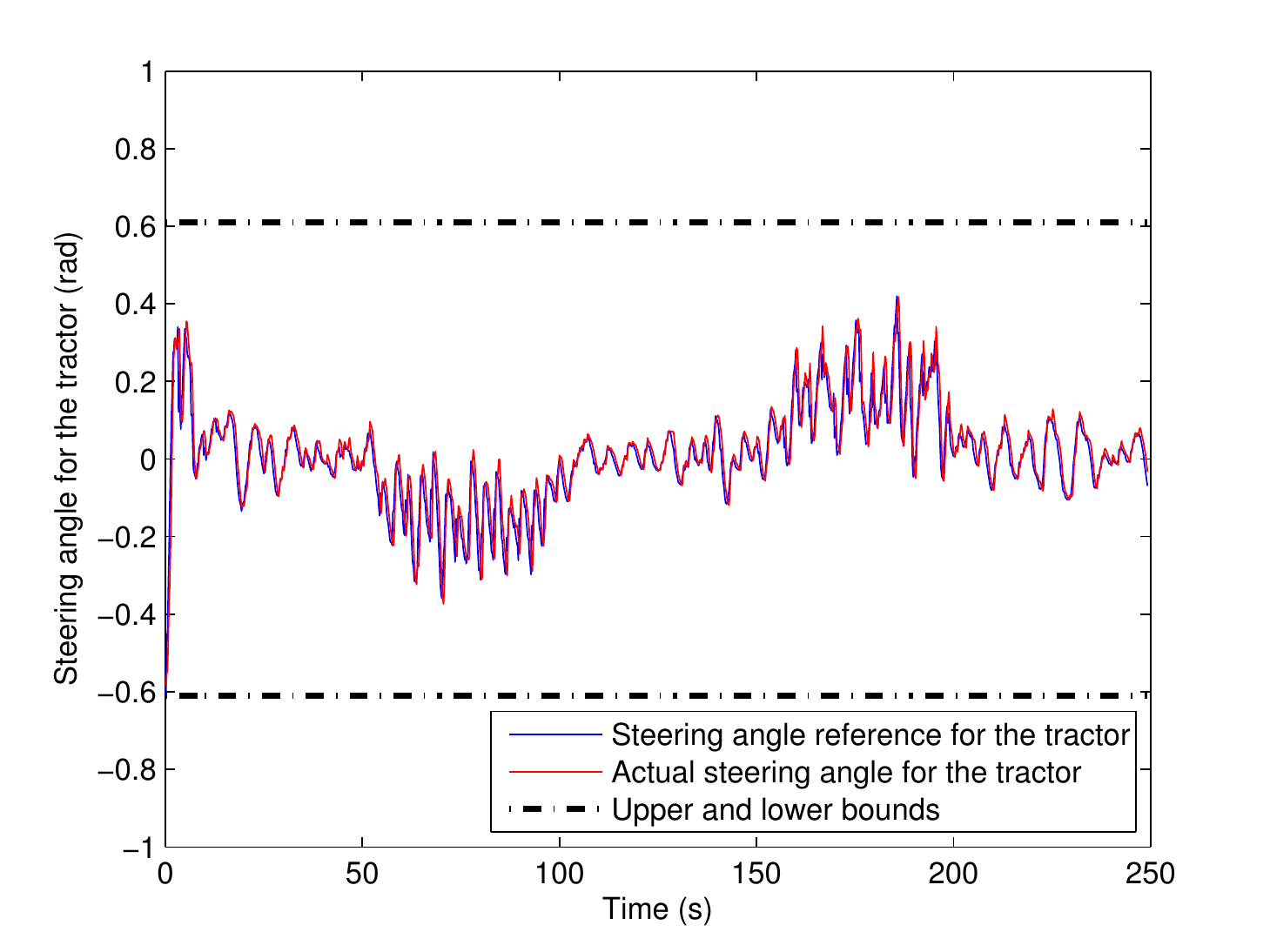}\\
\caption{Reference and actual steering angle for the tractor}
\label{steering_tra}
\end{figure}

\begin{figure}[h!]
\centering
\includegraphics[width=3.5in]{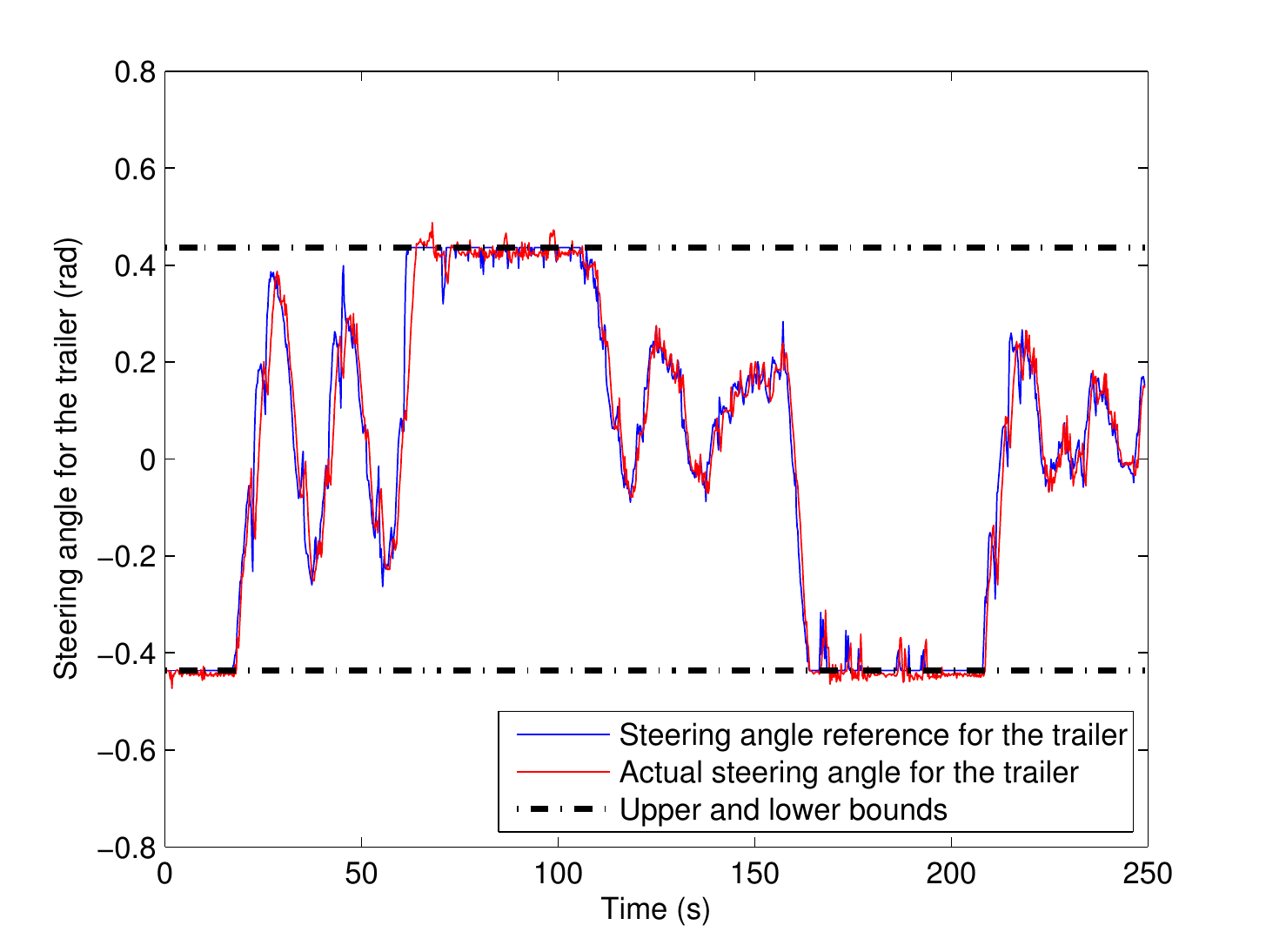}\\
\caption{Reference and actual steering angle for the trailer}
\label{steering_imp}
\end{figure}

In Figs. \ref{steering_tra}-\ref{steering_imp}, the outputs, the steering angle ($\delta^{t}$) reference for the tractor and the steering angle ($\delta^{i}$) reference for the trailer, of the DiNMPC are illustrated. The dashed lines indicate the physical constraints. As can be seen from these figures, the performance of the low level controllers is sufficient. Moreover, it is observed from Fig. \ref{steering_imp} that even if the output of the DiNMPC for the trailer reaches its constraints, the error correction is limited due to the aforementioned limited length of the drawbar.


\begin{figure}[h!]
\centering
\includegraphics[width=3.5in]{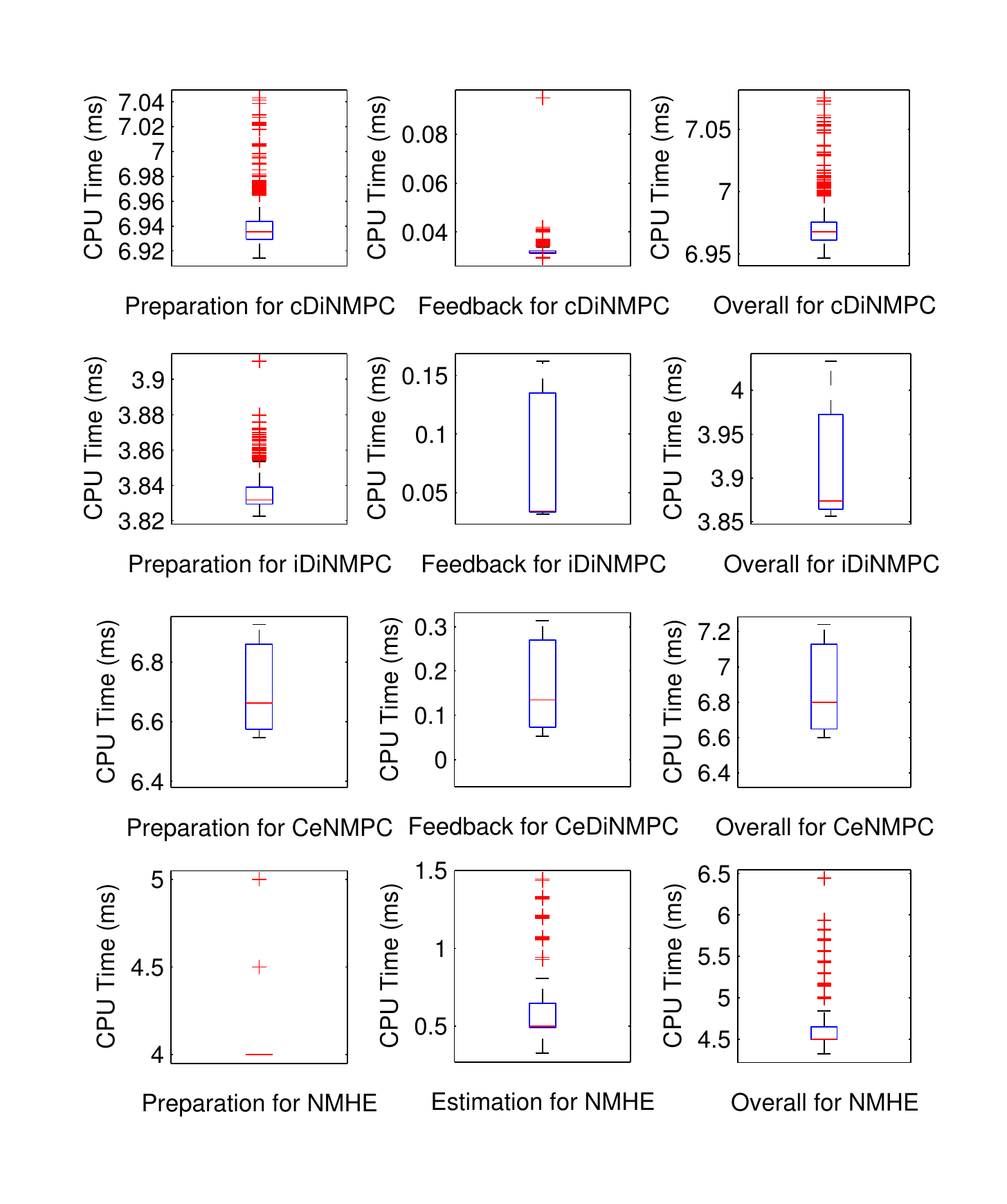}\\
\caption{Execution times of cDiNMPC, iDiNMPC, CeNMPC and NMHE.}
\label{mpcmheperformance}
\end{figure}

The execution times for cDiNMPC, iDiNMPC, Centalized NMPC (CeNMPC) and NMHE are given in Fig. \ref{mpcmheperformance}. During the real-time experiments, a real-time controller equipped with a 2.26 GHz Intel Core 2 Quad Q9100 quad-core processor (NI PXI-8110, National Instruments, Austin, TX, USA) has been used. The NMHE and NMPC routine have been assigned to one processor. As can be seen from this table, all the controllers are able to solve the optimization problems in $7$ ms when the \emph{ACADO} code generation tool is used. Since our system is naturally decoupled, the overall computation times for cDiNMPC and CeNMPC are similar, because they involve similar optimization problems in each case is similar to each other. On the other hand, since the optimization problem in iDiNMPC is relatively simpler than the one in CeNMPC and cDiNMPC, the computation time is considerably lower. Additionally, it is difficult to make a fair comparison between the mentioned different NMPC structures because of the relatively simple structure of the model. These differences in computational cost will become more critical when a more complex dynamic model is used.

It is to be noted that the decentralized case has less computational burden than the distributed case. However, the mean value of the Euclidian error to the straight line was found to be 7.95 cm and 5.42 cm for the tractor and trailer in the decentralized case, respectively. As can be seen from these results, the distributed case gives relatively better results than the decentralized one at the price of a limited increase in the computation time (around +300\% for iDINMPC and +600\% for cDiNMPC)

\section{Conclusions and future research} \label{Conc}
\subsection{Conclusions}
In this study, a fast DiNMPC-NMHE framework has been elaborated for the control of an autonomous tractor-trailer system. The state and the parameter estimation results show that NMHE is capable of accurately estimating the states and parameters while respecting the physical constraints.. The experimental results show that the proposed DiNMPC-NMHE framework is able to control the tractor-trailer system with a reasonable accuracy. The mean value of the Ecludian error to the straight line trajectory is $3.33$ cm and $3.22$ cm for the tractor and the trailer, respectively. It is to be noted that the iDiNMPC algorithm is only needed half the computation time required for CeNMPC and cDiNMPC, while similar tracking performance was obtained. So, it is a more efficient control strategy for autonomous tractor-trailer trajectory tracking. However, the cDiNMPC gives similar computation time to the CeNMPC due to the fact that our system is naturally decoupled.

\subsection{Future research}
Since the DiNMPC-NMHE framework based upon the adaptive kinematic model of the tractor-trailer system provides feedback times around $4-7$ ms, the design can be extended for a dynamic model. Whereas in case of having a kinematic model, a cDiNMPC and a iDiNMPC are designed for the tractor and the trailer subsystems, respectively, two cDiNMPCs can be designed for the case of having a dynamic model. The difference between the kinematic and the dynamic models is that while the input to the second subsystem (the trailer) affects the first subsystem (the tractor) for case of a dynamic model, it does not have such an effect for the case of an kinematic model.

\section*{Acknowledgment}
This work has been carried out within the IWT-SBO 80032 (LeCoPro) project funded by the Institute for the Promotion of Innovation through Science and Technology in Flanders (IWT-Vlaanderen). We would like to thank Mr. Soner Akpinar for his technical support for the preparation of the experimental set up.

\bibliographystyle{elsarticle-num}
\bibliography{references_file}

\end{document}